\documentstyle[preprint,aps]{revtex}

\begin{document}
\title {de Gennes-Saint-James resonant transport in Nb/GaAs/AlGaAs heterostructures}
\author{Francesco Giazotto$^{a)}$, Pasqualantonio Pingue, and Fabio Beltram}
\address{NEST-INFM and Scuola Normale Superiore, I-56126 Pisa, Italy}
\author {Marco Lazzarino, Daniela Orani, Silvia Rubini, and Alfonso Franciosi$^{b)}$ }
\address{Laboratorio Nazionale TASC-INFM, Basovizza, I-34012 Trieste, Italy}
\footnotetext[1]{Electronic mail: giazotto@nest.sns.it}
\footnotetext[2]{Also with Dipartimento di Fisica, Universit\`a di Trieste, I-34127 Trieste, Italy}
\maketitle

\begin{abstract}
Resonant transport is demonstrated in a hybrid superconductor-semiconductor heterostructure junction grown by molecular beam epitaxy on GaAs. This heterostructure realizes the model system introduced by de Gennes and Saint-James in 1963 [P. G. de Gennes and D. Saint-James, Phys. Lett. {\bf 4}, 151 (1963)]. At low temperatures a single marked resonance peak is shown superimposed to the characteristic Andreev-dominated subgap conductance. 
The observed magnetotransport properties are successfully analyzed within the random matrix theory of quantum transport, and ballistic effects are included by directly solving the Bogoliubov-de Gennes equations.
\end{abstract}
\vskip0.5cm
\newpage

Transport dynamics in mesoscopic devices comprising superconducting electrodes is a subject of increasing interest and rapid development thanks to the progress in nanofabrication techniques and materials science \cite{nanofab}. Transmission of quasi-particles through superconductor-normal metal (SN) interfaces requires conversion between dissipative currents and dissipationless supercurrents and is made possible by a two-particle process known as {\it Andreev reflection} (AR) \cite{andr}. 
An electron injected from  the normal metal  with energy lower than the superconductor gap is reflected as a phase-matched hole, while a Cooper pair is transmitted in the superconductor.
Due to its two-particle nature, AR is strongly affected by the transmissivity at the SN interface and much effort has to be devoted to the optimization of this parameter \cite{lach,kroemer,tak,tabor,giaz}.  
In the presence of scattering centers in the normal region, the phase relationship between incoming and retroreflected particles can give rise to marked coherent-transport phenomena such as {\it reflectionless tunneling} \cite{kast91,magnee,sanq,nbgaas}. 
One particularly interesting case is that of a single scatterer represented by an insulating barrier (I) inserted in the structure during growth in the normal region. This configuration can give rise to controlled interference effects. Among these, one of the most intriguing is represented by de Gennes-Saint-James  resonances \cite{GJ} in SNIN systems where the N interlayer is characterized by a constant pair potential. The case of a null pair potential is especially relevant to the present case. 
Multiple reflections off the superconductive gap (i.e. Andreev reflections) and off the insulating barrier (i.e. normal reflections) may give rise  to quasi-bound states \cite{GJ,bagwell1} that manifest themselves as conductance resonances. Transport resonances linked to similar multilayer configuration were observed experimentally in {\it all-metal} structures \cite{RM,wong,tessmer} and provided elegant evidence of quasi-particle coherent dynamics in SN systems.
 
In this Letter we report the experimental observation of de Gennes-Saint-James (dGSJ)-type resonances in a microstructure consisting of a Nb/GaAs/AlGaAs/GaAs hybrid heterojunction. This result was made possible by the exploitation of semiconductor epitaxial growth to tailor electronic states according to the original dGSJ model system. We analyzed the data within the random matrix theory of quantum transport and included ballistic effects by directly solving the Bogoliubov-de Gennes equations in a model potential profile. 
Our description of the system is confirmed by the observed temperature and magnetic-field dependence.

A sketch of the Nb/GaAs/AlGaAs structure is shown in Fig. 1(a). The semiconductor portion consists of a 1 $\mu$m thick $n$-GaAs(001) buffer layer Si-doped nominally at $n= 2\times 10^{18}$ cm$^{-3}$ grown by molecular beam epitaxy (MBE) on a $n$-GaAs(001) substrate, followed by a 4 nm thick Al$_{0.3}$Ga$_{0.7}$As barrier. This was followed by the growth of a 12 nm thick GaAs(001) epilayer Si-doped at $n= 2\times 10^{18}$ cm$^{-3}$ and by 14 nm of GaAs doped by a sequence of six Si $\delta$-doped layers spaced by 2 nm. A 1-$\mu$m-thick amorphous As cap layer was deposited in the MBE growth chamber to protect the surface during transfer in air to an UHV sputter-deposition/surface analysis system. 
After thermal desorption of the As protective cap layer, a 100-nm-thick Nb film was then deposited {\it in situ} by DC-magnetron sputtering.
The thickness of the GaAs epilayer sandwiched between the superconductor and the AlGaAs barrier was selected in order to have an experimentally-accessible single quasi-bound state below the superconductive gap, and the Si $\delta$-doped layers at the Nb/GaAs interface were employed to achieve the required transmissivity. More detail on the contact fabrication procedure is reported elsewhere \cite{nbgaas} where the reference SN junctions are also described. The latter consisted of a Nb/$\delta$-doped-GaAs junction {\it without} the AlGaAs insulating barrier. A qualitative sketch of the energy-band diagram of our SNIN structure is depicted in Fig. 1(b).

100$\times$160\,$\mu$m$^2$ junctions were defined by standard photolithographic techniques and reactive ion etching. 4-wire measurements were performed with two leads on the junction under study and the other two connected to the sample back contact.

Figure 2 shows the measured differential conductance $vs$ bias ($G(V)$) for the resonant structure (SNIN structure, panel (a)) and for the reference junction (SN structure, panel (b)) at $T=0.34$ K. Comparison of the two characteristics clearly shows the presence of a marked subgap conductance peak in the SNIN, resonant device. As we shall argue this is the first demonstration of dGSJ resonant transport in a hybrid superconductor-semiconductor system. The resonance is superimposed to the typical Andreev-dominated subgap conductance. The symmetry of conductance and the zero-bias conductance peak (ZBCP) peculiar to reflectionless tunneling (RT) further demonstrate the effectiveness of the fabrication protocol.

Quantitative determination of the resonant transport properties in such a system is not trivial as it can be inferred by inspecting Fig. 1(b) and considering the diffusive nature of the normal regions. dGSJ-enhancement, however, is an intrinsically ballistic phenomenon so that its essential features can be captured with relative ease. One study particularly relevant for our system was performed in Ref.\cite{bagwell1}. In the context of ballistic transport a one-dimensional SNIN structure was studied as a function of the N interlayer thickness $d$ and it was demonstrated that resonances can occur in the subgap conductance spectrum for suitable geometric conditions. The insulating barrier  was simulated by a $\delta$-like potential ${\mathcal V}(x)= {\mathcal V}{_I} \delta (x-d)$.  Customarily the barrier strength is described by the dimensionless coefficient $Z= {\mathcal V}{_I}/ \hbar v_F$, where $v_F$ is the electron Fermi velocity \cite{btk}, and in the following we shall make use of it to characterize our system. 
The key-results of the analysis are: ({\it a}) the number of resonances increases for larger thickness $d$ of the metallic interlayer and is virtually independent of $Z$; ({\it b}) the energy-width of such resonances decreases for increasing $Z$. 

The one-dimensional differential conductance $G(V)$ at temperature $T$ can be expressed as \cite{btk}
\begin{equation}
G(V) ={\mathcal G }\int_{-\infty}^{\infty}{\mathcal T(E)}\,f_{0}^{2}({\mathcal E} -qV) \, e^{\frac{{\mathcal E}-qV}{k_{B}T}}d{\mathcal E},
\label{conductance}
\end{equation}
where ${\mathcal G }=q^2/\pi k_{B}T\hbar$, ${\mathcal T(E)}=[1+A({\mathcal E})-B({\mathcal E})]$, and $A(\mathcal E)$ and $B(\mathcal E)$ are the energy-dependent Andreev  and normal reflection transmission probabilities for the complete SNIN structure, respectively. $qV$ is the incoming particle energy and $f_{0}({\mathcal E})$ is the Fermi distribution function.  
This model is rather idealized, but is a useful tool to grasp the essential features of our system such as number and position of resonances \cite{bagwell2}. 
In order to apply it we must first determine  parameters such as barrier strength, electron density and mean free path. Also, any quantitative comparison with experiment requires us to estimate the sample series resistance, which influences the experimental energy position of the resonance peak. The above parameters can be obtained from an analysis of the RT-driven ZBCP and from Hall measurements.

We performed Hall measurements at 1.5 K and obtained carrier density $n\simeq 4\times 10^{18}$\,cm$^{-3}$ and mobility $\mu \simeq 1.5\times 10^{3}$\,cm$^{2}$/V s. These data allow us to estimate the thermal coherence length $L_T =\sqrt{\hbar D/2\pi k_{B}T}=\,0.13\,\mu$m$/\sqrt{T}$, where $D=1.34\times 10^2$\,cm$^2$/s is the diffusion constant, and the electron mean free path ${\ell}_{m}\,\approx 48$\,nm.  $ {\ell}_{m}$ compares favorably with the geometrical constraints of the structure ($d=26$ nm $< {\ell}_{m}$) and further supports our ballistic analysis.

The ZBCP can be described following the work by Beenakker and co-workers  \cite{marmor}. Their model system consists of a normal disordered region  of length $L$ and width $W$ separated by a finite-transparency  barrier from a superconductor. Under appropriate conditions a ZBCP is predicted to occur, whose width $V_c$ (i.e., the voltage at which the ZBCP is suppressed) was estimated to be of the order of the Thouless voltage, $V_c =\hbar D/e L^2$. Similarly, upon application of a magnetic field, the ZBCP is suppressed for fields of the order of $B_c=h/eLW$. In real systems, at finite temperatures, $L$ and $W$ are to be replaced by $L_T$, if it is smaller \cite{marmor}. 
At 0.34 K $L_T \approx 0.22$\,$\mu$m and in our junctions we calculate $V_c \simeq 200$\,$\mu$V. Comparison with the experimental value $V_c^{exp} \simeq 600$\,$\mu$V in Fig. 3 (see solid line), allowed us to estimate the series resistance contribution to the measured conductivity. This rather large effect stems mainly from the AlGaAs barrier and the sample back-contact resistance.

Following Eq. \ref{conductance} we calculated the conductance for the nominal N-thickness value ($d = 26$ nm) and $Z = 1.4$, as appropriate for the AlGaAs barrier \cite{quantiz}. We emphasize, however, that the essential features such as the number and energy position of the dGSJ resonances are virtually independent of Z and are controlled instead by the value of $d$. Our calculations yield a single conductance peak at energies corresponding to about $0.8\,\Delta$, where $\Delta$ is the superconductor energy gap. By including the above-determined series resistance contribution, the resonance peak is positioned at about $3.5\,$mV (see Fig. 3, dashed line, $T=0.34$ K). 
The corresponding experimental value is about 3 mV (solid line in Fig. 3), but the observed energy difference is well within the uncertainty resulting from the determination of the series resistance.
The results of our model calculations strongly support our interpretation of the experimental structure in terms of dGSJ resonant transport.
Previous results on all-metal films generally showed more complex structures resulting from several dGSJ peaks, consistent  with the wider N-regions employed and the larger $k_F$ values characteristic of metallic systems. Indeed, our calculations show that for our material system configurations presenting more than one peak (i.e. more quasi-bound states) are not experimentally accessible. More peaks, in fact, would be present for interlayer thickness largely exceeding the required quasi-particle coherence length \cite{wolf}. The limited value of the latter also hinders a detailed study of the peak-position dependence on interlayer thickness.     
 
Ordinary resonant tunneling in the normal double-barrier potential
schematically shown in Fig. 1(b) cannot explain the observed subgap structure. This is indicated
by the symmetry in the experimental data for positive and negative bias and is further proven
by the temperature and magnetic field dependence of the differential conductance. 
Figure 4 shows a set of $G(V)$s measured in the 0.34-1.55 K range for the ZBCP, (a), and for the resonance peak, (b). Both effects show a strong dependence on temperature and at $T=1.55$ K are totally suppressed. This temperature value is within the range where RT suppression is expected \cite{kast91,magnee,sanq,nbgaas}.
At higher temperatures, the conductance in the resonance region resembles that of the reference SN junction of Fig. 2(b).
Notably, the ZBCP and the resonance peak disappear at the same temperature, hence indicating the coherent nature of the observed effect.  

Further confirmation of the nature of the resonance peak can be gained observing its dependence from the magnetic field. Figure 5 shows $G(V)$ at $T=0.35$\,K for several values of the magnetic field applied in the plane of the junction for the ZBCP, (a), and for the resonance peak, (b). 
The measurements confirm the known sensitivity of ZBCP to the magnetic field \cite{marmor}, and clearly indicate that the dGSJ resonance is easily suppressed for critical fields of the order of 100 mT. 

We also investigated the perpendicular field configuration and also in this case 
 the resonance and the ZBCP displayed a similar dependence (data not shown). Such behavior is fully consistent with dGSJ-related origin but is not compatible with a normal resonant tunneling description of the data \cite{rtsemicond}.  

In summary, we have experimentally observed dGSJ resonant states in Nb/GaAs/AlGaAs hybrid heterostructures. 
Transport was studied as a function of temperature and magnetic field and was successfully described within the ballistic model of Riedel and Bagwell\cite{bagwell1}.
To the best of our knowledge this result represents the first demonstration of  dGSJ resonant transport in superconductor-semiconductor hybrid structures and was made possible by the fabrication procedure adopted in this work.
The present results suggest that the Nb/GaAs/AlGaAs system is a good candidate for the implementation of complex mesoscopic structures that can take advantage of the mature AlGaAs nanofabrication technology. Such structures may represent ideal prototype systems for the study of coherent transport and the implementation of novel hybrid devices. 

This work was supported by INFM under the PAIS project Eterostrutture Ibride Semiconduttore-Superconduttore.



\begin{figure}
\caption{ (a) Schematic structure of the Nb/GaAs/AlGaAs system  analyzed in this work. (b) Sketch of the  energy-band diagram of the structure. The shaded area represents de Gennes-Saint-James quasi-bound state confined between the superconductor and the Al$_{0.3}$Ga$_{0.7}$As barrier.}
\label{F1}
\end{figure}

\begin{figure}
\caption{ Differential conductance $vs$ voltage at $T=0.34$ K for the SNIN junction, (a), and for the reference SN junction, (b). Both curves show  reflectionless tunneling enhancement around zero bias, while only in (a) a finite-bias, subgap de Gennes-Saint-James peak is present.}
\label{F2}
\end{figure}

\begin{figure}
\caption{ Comparison at $T=0.34$ K between the experimental differential conductance $G(V)$ (solid line) and the numerical simulation from Eq. 1 (dotted line, see text).}
\label{F3}
\end{figure}

\begin{figure}
\caption{ Differential conductance $vs$ voltage at several temperatures in the 0.34 to 1.55\,K range (a) for the zero bias conductance peak, and (b) for the de Gennes-Saint-James resonance peak. 
Data were taken at: (a) $T=0.34,0.45,0.55,0.65,0.75,0.85,0.95,1.05,1.26,1.45,1.50,1.55$ K; (b) $T=0.34,0.55,0.65,0.85,1.05,1.26,1.55$ K.}
\label{F4}
\end{figure}

\begin{figure}
\caption{ Differential conductance $vs$ voltage at $T=0.35$\,K for several values of the magnetic field applied in the plane of the junction. Dependence (a) of the  zero bias conductance, and (b) of the de Gennes-Saint-James resonance peak. In (b) data were taken at $H_{//}=0,10,20,25,30,35,40,45,50,60,70,80,100,150$ mT.}
\label{F5}
\end{figure}

\begin{references}

\bibitem{nanofab} C. J. Lambert and R. Raimondi, J. Phys.: Condens. Matter {\bf 10}, 901 (1998).

\bibitem{andr} A. F. Andreev, Zh. Eksp. Teor. Fiz. {\bf 46}, 1823 (1964)
	[Sov.Phys. JETP {\bf 19}, 1228 (1964)].

\bibitem{lach} S. G. Lachenmann {\it et al.}, J. Appl. Phys. {\bf 83}, 8077 (1998).

 \bibitem{kroemer} C. Nguyen {\it et al.}, Appl. Phys. Lett. {\bf 65}, 103 (1994).

\bibitem{tak} T. Akazaki {\it et al.}, Appl. Phys. Lett. {\bf 59}, 2037 (1991).

\bibitem{tabor} R. Taboryski {\it et al.}, Appl. Phys. Lett. {\bf 69}, 656 (1996).

\bibitem{giaz} S. De Franceschi {\it et al.}, Appl. Phys. Lett. {\bf 73}, 3890 (1998).

\bibitem{kast91} A. Kastalsky {\it et al.}, Phys. Rev. Lett. {\bf 67}, 3026
	(1991).

\bibitem{magnee}  P. H. C. Magn\'ee {\it et al.}, Phys. Rev. B {\bf 50}, 4594 (1994).

\bibitem{sanq} W. Poirier {\it et al.}, Phys. Rev. Lett. {\bf 79}, 2105 (1997).

\bibitem{nbgaas} F. Giazotto {\it et al.}, Appl. Phys. Lett {\bf 78}, 1772 (2001).

\bibitem{GJ} P. G. de Gennes and D. Saint-James, Phys. Lett. {\bf 4}, 151 (1963).

\bibitem{bagwell1} R. A. Riedel and P. F. Bagwell, Phys. Rev. B {\bf 48}, 15198 (1993).

\bibitem{RM} J. M. Rowell, Phys. Rev. Lett. {\bf 30}, 167 (1973).

\bibitem{wong} L. Wong {\it et al.}, Phys. Rev. B {\bf 23}, 5775 (1981).

\bibitem{tessmer} S. H. Tessmer {\it et al.}, Phys. Rev. Lett. {\bf 70}, 3135 (1993).

\bibitem{btk} G. E. Blonder {\it et al.}, Phys. Rev. B {\bf 25}, 4515 (1982).

\bibitem{bagwell2} We confined our calculation to a one-dimensional case in light of the results by S. Chaudhuri and P. F. Bagwell, Phys. Rev. B {\bf 51}, 16936 (1995). These authors showed the insensitivity to dimensionality of the essential properties of transport resonances. Resonances are determined by inspecting the $G(V)$ behavior.

\bibitem{marmor} I. K. Marmorkos {\it et al.}, Phys. Rev. B {\bf 48}, 2811 (1993); C. W. J. Beenakker {\it et al.}, Phys. Rev. Lett.	{\bf 72}, 2470 (1994).

\bibitem{quantiz} Taking into account the Al$_{0.3}$Ga$_{0.7}$As/GaAs discontinuity, Fermi energy and barrier thickness it is straightforward to determine $Z\simeq 1.4$.

\bibitem{wolf} See E. L. Wolf, {\it Principles of Electron Tunneling Spectroscopy } (Oxford University Press, New York, 1985), p. 193.

\bibitem{rtsemicond} F. Capasso, {\it Physics of Quantum Electron Devices} (Springer, Berlin, 1990). 



\end{references}
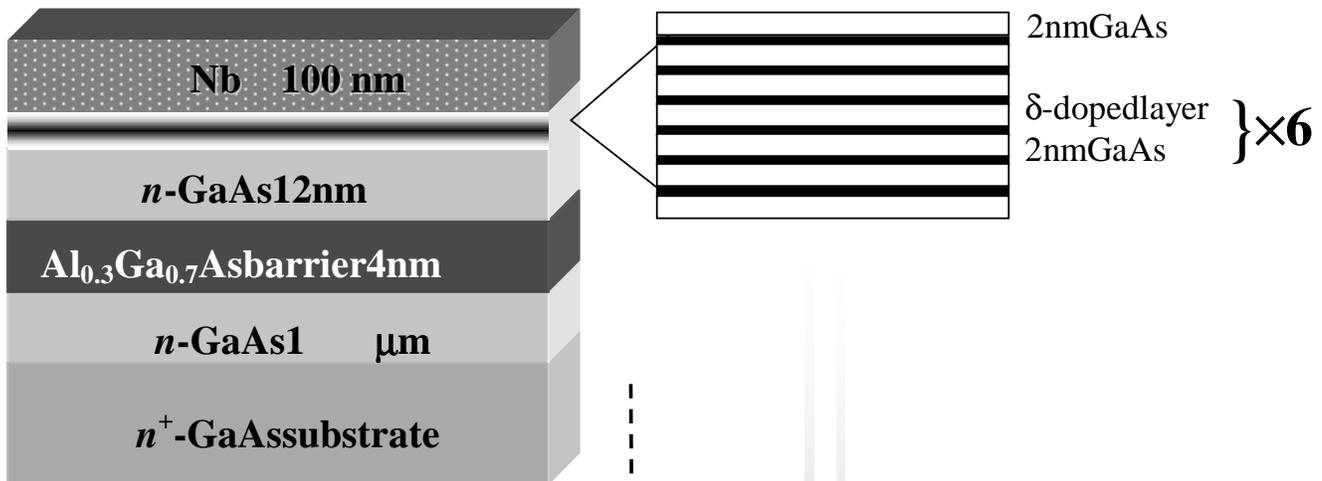
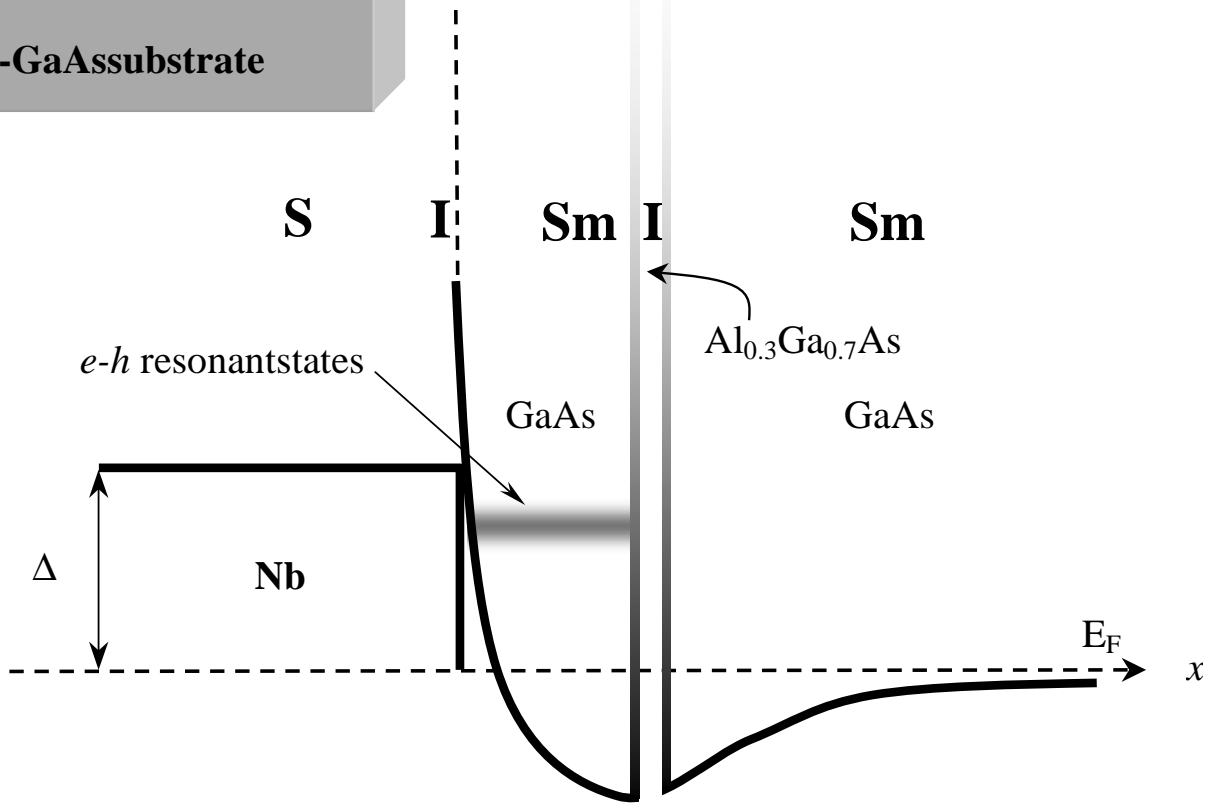
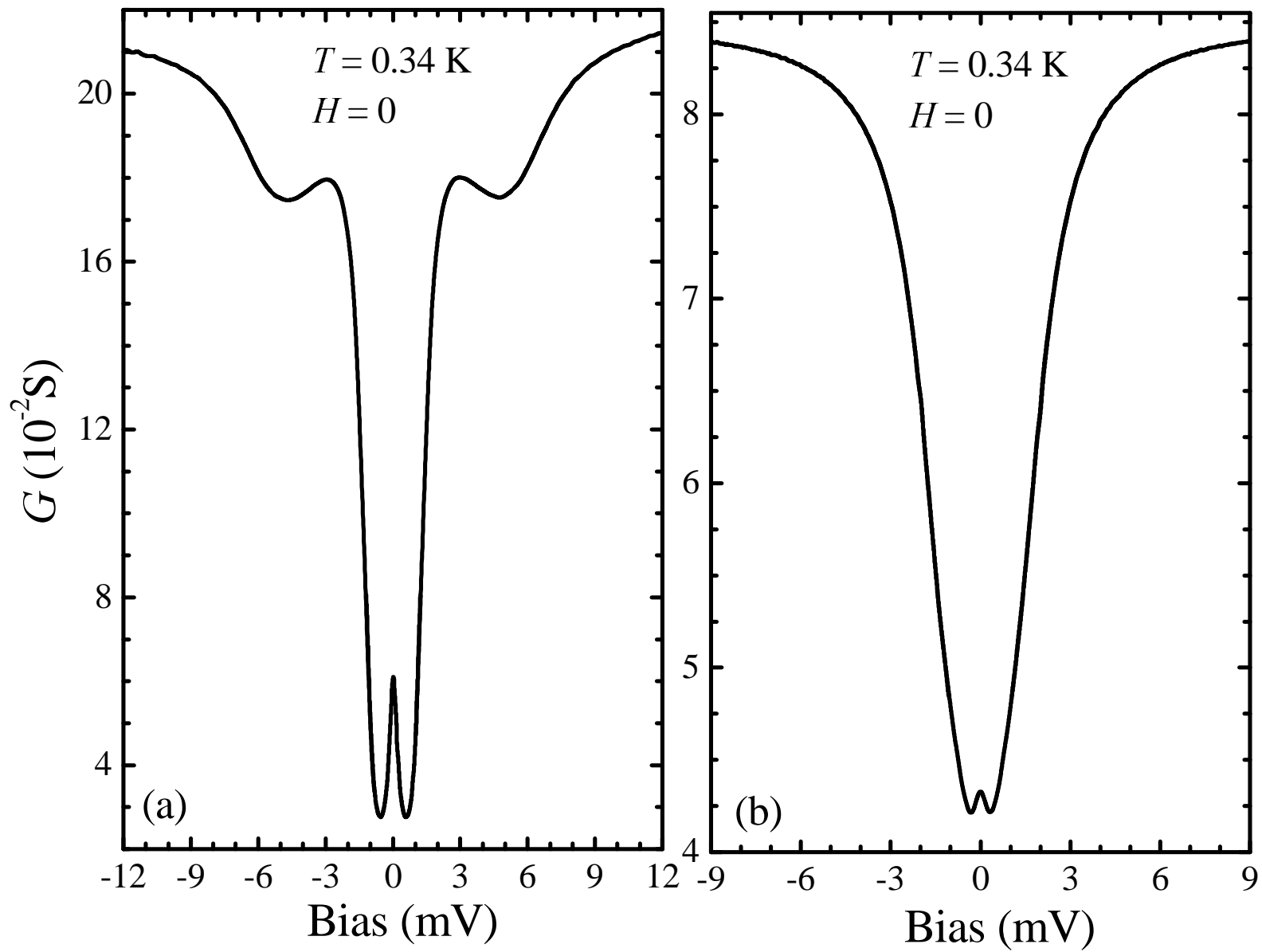
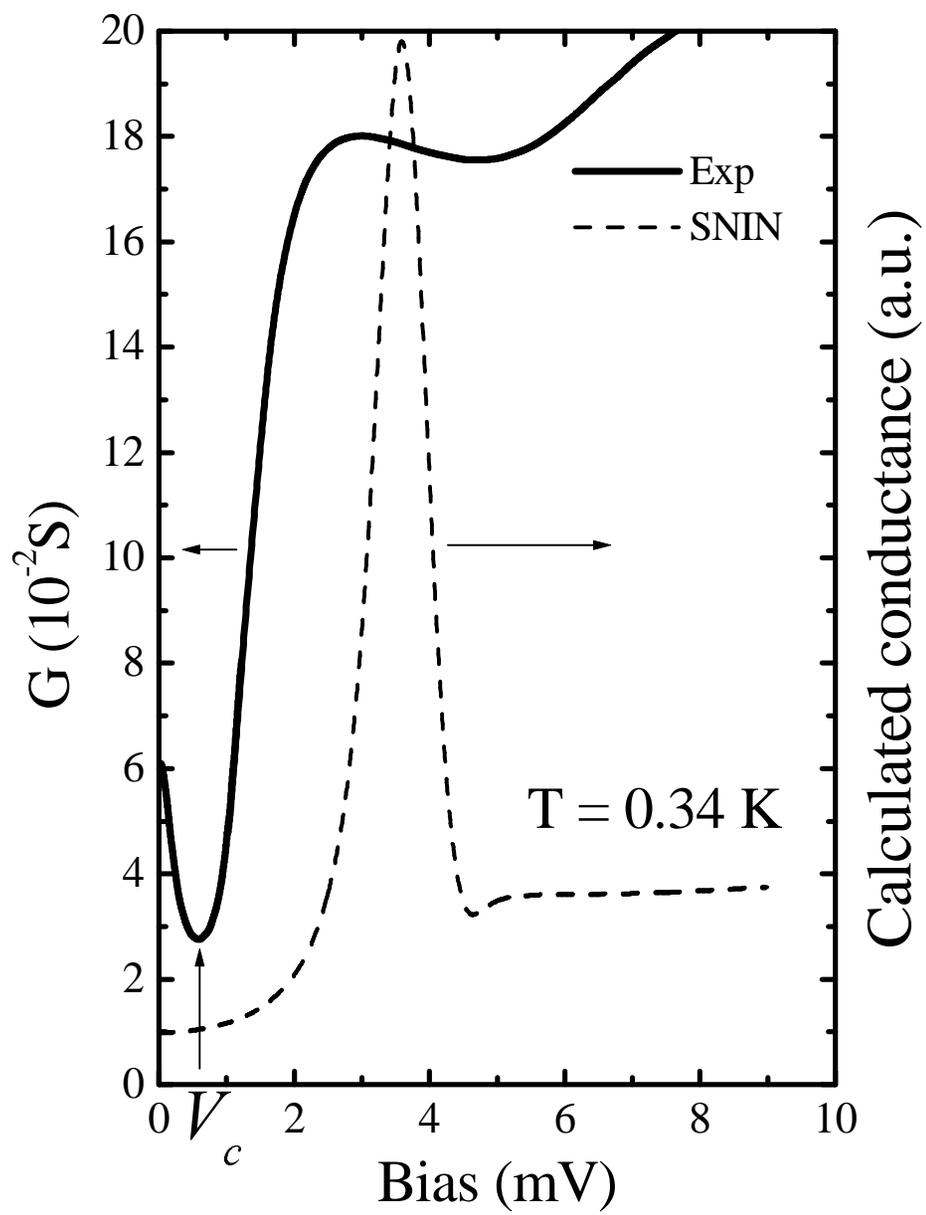
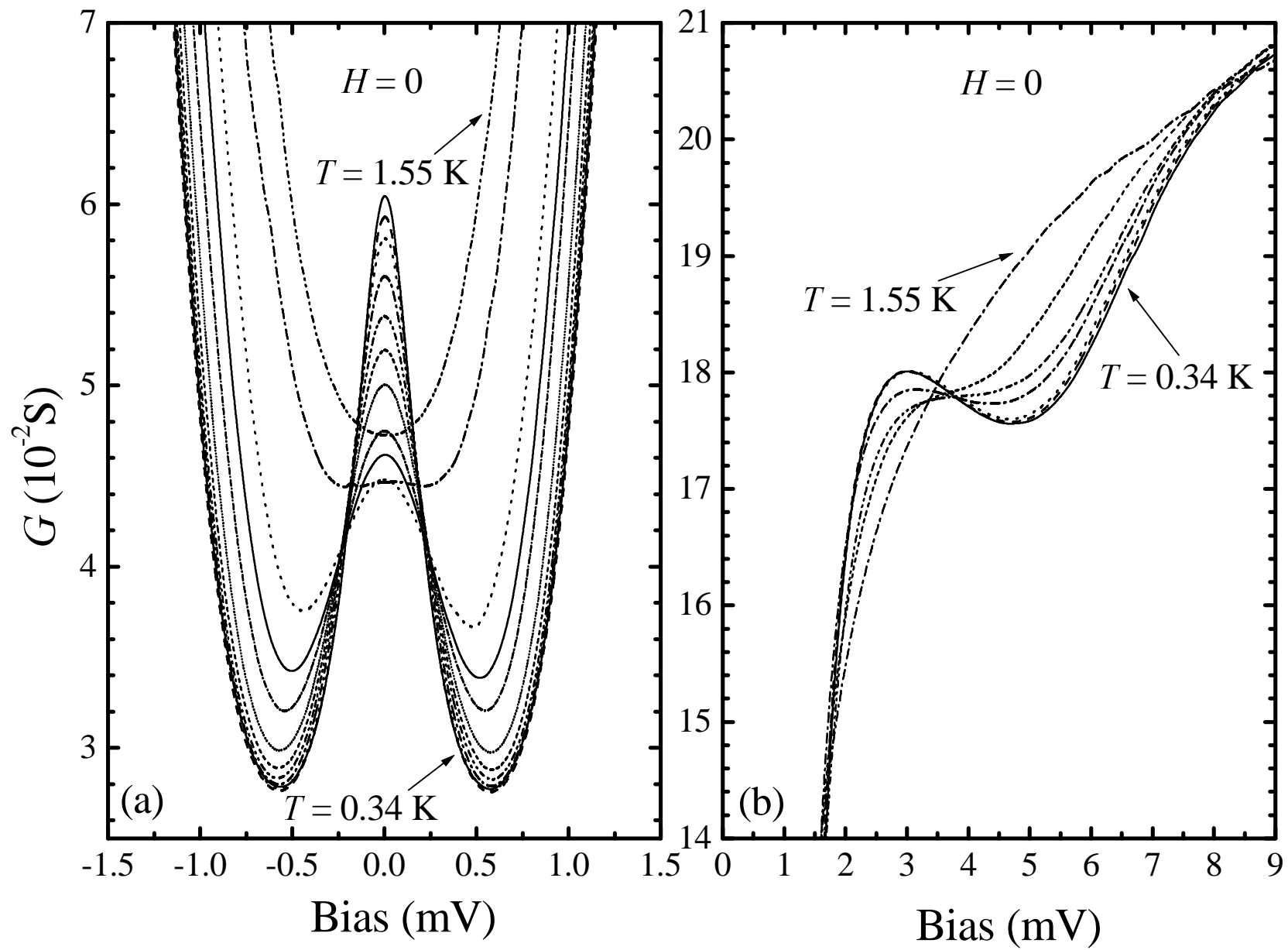
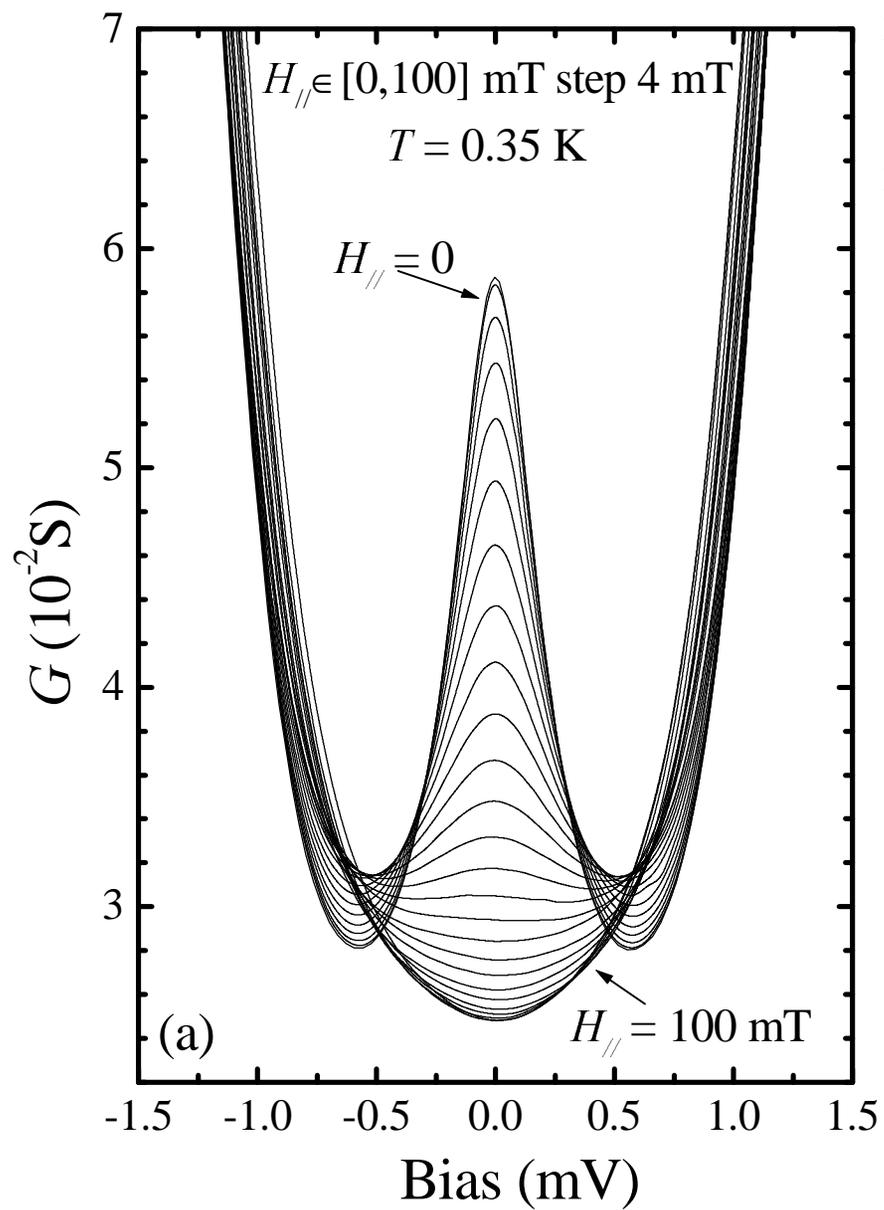
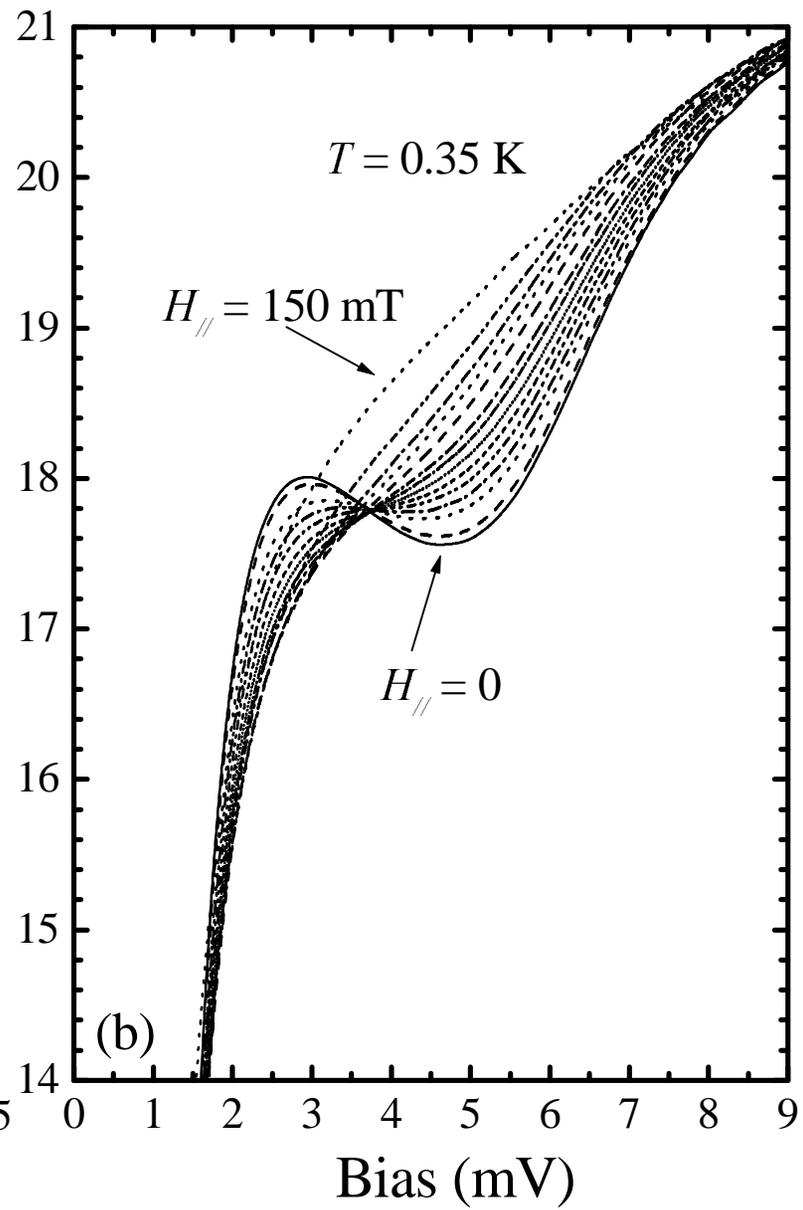
\end{document}